\documentclass[12pt,a4paper]{article}
\input xy
\xyoption{all}
\usepackage{amsthm}
\usepackage{amsfonts,amsmath}
\usepackage{lscape}
\usepackage{amsfonts,latexsym}
\usepackage{pstricks}
\usepackage{pst-grad}
\input epsf

\def\tp{\otimes}

\def\d{\delta}
\def\dn{\bar{\delta}}
\def\mod{\hspace{0.2cm} \mbox{mod} \hspace{0.2cm}} 

\def\N{\mathbb{N}}
\def\Z{\mathbb{Z}}

\def\R{\mathbb{R}}
\def\C{\mathbb{C}}

\def\L{\mathcal{L}}

\newtheorem{thm}{Theorem}[section]
\newtheorem{prop}[thm]{Proposition}

\newtheorem{obs}[thm]{Observation}

\theoremstyle{definition}

\theoremstyle{remark}

\title{Ground-state phase diagram for a system of interacting, $D(D_3)$ non-Abelian anyons}
\author{P.E. Finch$^1$, H. Frahm$^1$, and J. Links$^2$ \\
  \\
  $^1$Institut f\"ur Theoretische Physik, \\
  Leibniz Universit\"at Hannover, \\
  Appelstra\ss e 2, 30167 Hannover, Germany \\
  \\
 $^2$Centre for Mathematical Physics,\\
  School of Mathematics and Physics, \\ 
 The University of Queensland, 4072, Australia}
\date{}

\begin{document}
\maketitle

\begin{abstract}

We study an exactly solvable model of $D(D_3)$ non-Abelian anyons on a one-dimensional lattice with a free coupling parameter in the Hamiltonian.  For certain values of the coupling parameter level crossings occur, which divide the ground-state phase diagram into four regions. We obtain explicit expressions for the ground-state energy in each phase, for both closed and open chain boundary conditions. For the closed chain case we show that chiral phases occur which are characterised by non-zero ground-state momentum.  
\end{abstract}

\section{Introduction}

Many-body non-Abelian anyonic states are characterised by non-trivial transformation upon particle interchange, a property called {\it braiding} which underpins the theory of topological quantum computation \cite{nssfd2008}. For consistency, braiding transformations must be compatible with a set of fusion rules which govern the decomposition of product states in a manner that preserves the required symmetries. These compatibility requirements are automatically met if the symmetry algebra has the stucture of a quasi-triangular Hopf algebra \cite{kitaev2003}.  A familiar example of a fusion rule arises in the case of spin-1/2 particles, whereby the state space for two spin-1/2 particles decomposes into triplet and singlet sectors. The well-known Heisenberg spin chain is a model that assigns different energies to the triplet and singlet sectors for neighbouring spins on a one-dimensional lattice. Analogously, one-dimensional models for  non-Abelian anyons with nearest-neighbour interactions may be formulated by assigning energies to the different nearest-neighbour sectors determined by the fusion rules. For the case of Fibonacci anyons, a detailed description of this approach can be found in \cite{ttwl08}.  Along similar lines there have been several recent studies of systems involving interacting non-Abelian anyons in order to better understand the properties of collective states in many-body systems \cite{by07,ftltkwf2007,frbm08,gatltw2009,tafhlt2008}.

One of the important features of the Heisenberg model is that it admits an
exact solution, as determined by Bethe in 1931 \cite{Bethe31}. In modern
approaches the exact solution may be obtained using the techniques of the
Yang-Baxter equation the Quantum Inverse Scattering Method (QISM) \cite{VladB}. Through this framework an advanced understanding of the Heisenberg model (and related models such as its anisotropic generalisation the $XXZ$ chain) continues to be developed in areas such as thermodynamics \cite{k04}, correlation
functions \cite{bjjst06,kkmst09}, and dynamics \cite{mc10,pschmwa06}.  Our objective is to adapt this general
program to derive a non-Abelian anyon system which is exactly solvable. Once
this model is obtained, the specific goal is to determine the quantum phase
transitions exhibited by the model and to investigate properties of the 
ground-state phases.

The symmetry algebra we employ is obtained through the Drinfeld double
construction \cite{d1986} applied to the group algebra of the dihedral group
$D_3$. Application of the double construction yields an algebra denoted
$D(D_3)$, which is necessarily quasi-triangular and consequently applicable as
a symmetry algebra for non-Abelian anyons. The algebra $D(D_3)$ has a finite
number of irreducible representations where the dimensions are either one, two
or three. Associated with a three-dimensional representation we present a
solution of the Yang-Baxter equation, apply the Quantum Inverse Scattering
Method to determine the Hamiltonian, and solve this model through Bethe ansatz
techniques. The Hamiltonian obtained in this manner has the feature that it can be expressed
as a linear combination of two mutually commuting terms. This property leads
to energy level crossings and as a result there are first-order quantum phase
transitions at which the ground-state energy has discontinuous first
derivative. Utilising the exact Bethe ansatz solution of the Hamiltonian we study the
ground-state phases. Our analysis begins for systems with closed boundary
conditions and we find that both time-reversal invariant and non time-reversal
invariant phases exist, which are in evidence by computing the momentum of the
ground state. We discover an unusual property at the
boundaries at which the ground-state energy level crossings occur, whereby the
degeneracy scales exponentially with system size. We also study the case of
open boundaries which, due to the lack of translational invariance, 
means that states are not characterised by momenta.
For this case we still find that the
ground-state phase boundaries due to level crossing are present, again with
exponential scaling of the degeneracy.
          
In section 2 we briefly recall some basic properties of the algebra $D(D_3)$. In section 3 we use a solution of the Yang--Baxter equation to derive an exactly solvable Hamiltonian for three different types of boundary conditions, which is solved through Bethe ansatz methods. To gain an understanding of the solutions of the Bethe ansatz equations which correspond to the ground state, we undertake numerical studies in section 4 and make several observations about the properties of these solutions. Guided by these we calculate the ground-state energy and momentum in section 5, and summarise our findings in section 6.

\section{Preliminaries}
We make use of the following two delta functions:
\begin{eqnarray*}
\d^j_i = 
	\begin{cases}
		1, & ~~ i = j, \\
		0, & ~~ i \neq j,
	\end{cases}
& \hspace{1cm} &
\dn_i^j = 
	\begin{cases}
		1, & ~~ i \equiv j \mod 3, \\
		0, & ~~ i \not\equiv j \mod 3,
	\end{cases}
\end{eqnarray*}
noting that the first delta function is not restricted to integer indices and at times will be used where the indices refer to elements of a group. \\

\noindent
We define $e_{i,j} \in M_{d \times d}(\C)$ to be the matrix with a one in the $i$th row and $j$th column and zeros elsewhere for $1 \leq i,j \leq d$. We extend these matrices by considering the indices modulo $d$, and they satisfy the relation
$$ e_{i,j}e_{k,l} = 
	\begin{cases}
		e_{i,l}, & ~~ j \equiv k \mod d, \\
		0, & ~~ j \not\equiv k \mod d,
	\end{cases} \hspace{1cm} i,j,k,l \in \Z.
$$ \\

\noindent
The algebra we will employ is based upon the dihedral group of order 6, $D_{3}$. This is the group of symmetries of a triangle, generated by two elements $\sigma$ and $\tau$, with the presentation
$$ D_{3} = \{\sigma, \tau | \sigma^{3} = \tau^{2} = \sigma \tau \sigma \tau = e \}, $$
where $e$ is the identity element of the group. Equivalently we could consider $S_{3}$, the permutation group on 3 objects, as it is isomorphic to $D_{3}$. The Drinfeld double \cite{d1986} of $D_{3}$ is the vector space:
$$ D(D_{3}) = \C\{gh^{*} | g,h \in D_{3}\} $$
where $\{h^*| h\in D_3\}$ is a basis for the dual algebra of $D_3$. Here we follow the notational conventions of \cite{cdil2010,DIL2006}.
Multiplication and comultiplication are defined by
$$ g_{1}h_{1}^{*} g_{2}h_{2}^{*} =  \d_{(h_{1}g_{2})}^{(g_{2}h_{2})} \, (g_{1}g_{2}) h_{2}^{*} \hspace{0.5cm} \mbox{and} \hspace{0.5cm} \Delta(gh^{*}) = \sum_{k\in D_{3}} g (k^{-1}h)^{*} \tp g k^{*}. $$
Details about the representation theory of $D(D_3)$, regarding construction of irreducible representations and decomposition of tensor products, can be found in \cite{DIL2006,FDIL2010}. We mention that there are only eight irreducible representations, two of dimension one, four of dimension two, and two of dimension three. We will be concerned with an anyonic chain for a particular three-dimensional local state space, and consequently we only present the associated representation of $D(D_{3})$ for this case:
\begin{eqnarray} 
\pi(\sigma) = \left(\begin{array}{ccc} 0 & 0 & 1 \\ 1 & 0 & 0 \\ 0 & 1 & 0 \end{array} \right), \hspace{0.5cm}
\pi(\tau) = \left(\begin{array}{ccc} 1 & 0 & 0 \\ 0 & 0 & 1 \\ 0 & 1 & 0 \end{array} \right), \hspace{0.5cm}
\pi(g^{*}) = \left(\begin{array}{ccc} \d_{g}^{\tau} & 0 & 0 \\ 0 & \d_{g}^{\sigma^{2}\tau} & 0 \\ 0 & 0 & \d_{g}^{\sigma\tau} \end{array} \right) .
\label{rep}
\end{eqnarray}

\section{The $R$-matrix and integrable models}
The zero-field six-vertex model $R$-matrix is given by
$$ r(z) = 
	\left(
	\begin{array}{cccc}
		w^{-1}z^{-1} - wz & 0 & 0 & 0\\
		0 & z^{-1} - z & w^{-1} - w & 0\\
		0 & w^{-1} - w & z^{-1} - z & 0\\
		0 & 0 & 0 & w^{-1}z^{-1} - wz
	\end{array}
\right), $$
where we will set $w=\exp({{2i\pi}/{3}})$.  A descendant of the zero-field six-vertex $R$-matrix is the following two-parameter $R$-matrix
\begin{eqnarray}
R(z_{1},z_{2}) = N(z_{1},z_{2})\sum_{a,i,j=1}^{n} \left[\sum_{b=1}^{n} w^{2a(b+j)} \overline{W}(z_{1}|b) \overline{W}(z_{2}^{-1}|b-j) \right] e_{i+j,i+a} \tp e_{i+a+j,i}, 
\label{rm}
\end{eqnarray}
where
$$ \overline{W}(z|l) = \left[\frac{z -  1}{wz  - w^{2}}\right]^{1-\dn_{l}^{0}} \hspace{0.7cm} \mbox{and} \hspace{0.7cm} N(z_{1},z_{2}) = -\frac{1}{3}(w z_{1} - w^{2}) (w - w^{2}z_{2}). $$
This two-parameter $R$-matrix is a special case of the Fateev--Zamolodchikov $R$-matrix \cite{fz82}, which itself is a special case of the chiral Potts $R$-matrix \cite{bpa88}. One can also show that it has the symmetry of $D(D_3)$ in the sense that 
$$ R(z_{1},z_{2}) \Delta(gh^{*})= {\Delta}^{T}(gh^{*}) R(z_{1},z_{2}), $$
where 
$$ {\Delta}^T(gh^{*}) = \sum_{k\in D_{3}} g k^{*} \tp g (k^{-1}h)^{*} $$
is the opposite comultiplication  
and the elements $gh^{*}$ are evaluated in the representation (\ref{rep}). Setting $z_1=z_2$ in (\ref{rm}) yields an $R$-matrix previously discussed in \cite{DIL2006}.

The above $R$-matrices are connected through the $L$-operator
$$ L(z) = \sum_{i=1}^{3}\left \{ \left(w^{i-1}e_{1,2} + w^{1-i}e_{2,1}\right) \tp e_{i,i} + z \left[ e_{1,1} \tp e_{i-1,i} + e_{2,2} \tp e_{i+1,i} \right] \right\}. $$
Altogether we have that these operators satisfy the four Yang-Baxter equations:
\begin{equation}
\label{eqnYBE}
\begin{aligned}
	r_{12}(x)r_{13}(xy)r_{23}(y) & = r_{23}(y)r_{13}(xy)r_{12}(x), \\
	r_{12}(x)L_{13}(xy)L_{23}(y) & = L_{23}(y)L_{13}(xy)r_{12}(x), \\
	L_{12}(x_{1})L_{13}(x_{1}y_{1})R_{23}(y_{1},y_{2}) & = R_{23}(y_{1},y_{2})L_{13}(x_{1}y_{1})L_{12}(x_{1}), \\
	R_{12}(x_{1},x_{2})R_{13}(x_{1}y_{1},x_{2}y_{2})R_{23}(y_{1},y_{2}) & = R_{23}(y_{1},y_{2})R_{13}(x_{1}y_{1},x_{2}y_{2})R_{12}(x_{1},x_{2}). \\
\end{aligned}
\end{equation}
These equations are identities on the tensor product of three spaces
$V_1\otimes V_2 \otimes V_3$, the subscripts of the $R$-matrices and
$L$-operators indicating the pair of spaces on which the objects act
non-trivially.
Additionally we have the properties
$$ R(1,1) = \Pi, \hspace{0.7cm} \Pi R(z_{1},z_{2}) \Pi = z_{1}z_{2} R(z_{2}^{-1},z_{1}^{-1}) \hspace{0.7cm} \mbox{and} \hspace{0.7cm} \left[R(z_{1},z_{2})\right]^{*} = R(z_{2}^{*},z_{1}^{*}), $$
where $\Pi$ is the usual permutation operator and $*$, now and hereafter, denotes complex conjugation.
These properties imply the additional
equation
$$ L_{12}^{*}(x_{2})L_{13}^{*}(x_{2}y_{2})R_{23}(y_{1},y_{2}) =
R_{23}(y_{1},y_{2})L_{13}^{*}(x_{2}y_{2})L_{12}^{*}(x_{2}). $$ 
\noindent
Using these operators we next construct integrable chains for three different
types of boundary conditions. \\

\noindent
\underline{\textbf{Periodic boundary conditions}}\\
\noindent
The most common application of the QISM is for the construction of closed
chains of $\L$ sites with periodic boundary conditions \cite{VladB}.  For
these systems we define the transfer matrices
$$ t^{(2,p)}(z) = \mbox{tr}_{0}\left[ L_{0\L}(z)...L_{01}(z) \right]
\hspace{0.7cm} \mbox{and} \hspace{0.7cm} t^{(3,p)}(z_{1},z_{2}) =
\mbox{tr}_{0}\left[ R_{0\L}(z_{1},z_{2})...R_{01}(z_{1},z_{2}) \right], $$
where the traces are taken over the auxiliary space $V_0$ (note that $\dim
V_0=2$ ($3$) for $t^{(2,p)}$ ($t^{(3,p)}$)).
As a consequence of the Yang-Baxter equations (\ref{eqnYBE}) these transfer
matrices commute for different values of the spectral parameters $z$, $z_1$,
$z_2$, i.e.
\begin{equation*}
\begin{aligned}
	\left[ t^{(2,p)}(x), t^{(2,p)}(y) \right] & =  0, \\
	\left[ t^{(2,p)}(x), t^{(3,p)}(y_{1},y_{2}) \right] & =  0, \\
	\left[ t^{(3,p)}(x_{1},x_{2}), t^{(3,p)}(y_{1},y_{2}) \right] & =  0\,.
\end{aligned}
\end{equation*}
These relations imply that the transfer matrices can be simultaneously
diagonalised and have eigenstates which are independent of the spectral
parameter.  This fact will be used for the construction of integrable
Hamiltonians below.
In addition, the transfer matrices satisfy the commutation relations
\begin{equation*}
\begin{aligned}
	\left[ \left(t^{(2,p)}(x)\right)^{*}, t^{(2,p)}(y) \right] & = 0, \\
	\left[ \left(t^{(2,p)}(x)\right)^{*}, t^{(3,p)}(y_{1},y_{2}) \right] &
        = 0.
\end{aligned}
\end{equation*}
Furthermore, as a consequence of $\left[t^{(2,p)}(z)\right]^{\dagger} =
t^{(2,p)}(z)$ for real $z$, where $\dagger$ denotes the hermitean conjugate, the spectra of $t^{(2,p)}(z)$ and
$\left(t^{(2,p)}(z)\right)^{*}$ are the same.  Finally, we note that by
construction  $t^{(2,p)}(z)$ is a polynomial of degree $\L$ in $z$, while 
the degree of each variable in  $t^{(3,p)}(z_{1},z_{2})$ is $\L$.

Within the family of commuting operators generated by the transfer matrix
$t^{(3,p)}(z_{1},z_{2})$ integrable Hamiltonians with local interactions can be constructed
as a consequence of the observation that $T=t^{(3,p)}(1,1)$ is the unitary
operator describing translations by one site, i.e. $T \mathcal{O}_i T^{-1} =
\mathcal{O}_{i+1}$ for operators $\mathcal{O}_i$ acting nontrivially on site
$i$ only.  $T^\L=1$ as a consequence of the periodic boundary conditions.
From the first order in a Taylor series of $\ln t^{(3,p)}(z_1,z_2)$ around
$z_1=z_2=1$ we obtain a lattice Hamiltonian with nearest neighbour
interactions:
\begin{equation}
\begin{aligned}
 \label{eqndefglobalperiodicHam}
  \mathcal{H} & = i\left\{ \alpha_{1} \left[
      \frac{\partial}{\partial z_{1}}\ln\left(t^{(3,p)}(z_{1},z_{2})\right) -
      \beta_{1} \L  \right] - \alpha_{2} \left[
      \frac{\partial}{\partial z_{2}}\ln\left(t^{(3,p)}(z_{1},z_{2})\right) -
      \beta_{2} \L \right] \right\}_{z_{1}=1,z_{2}=1}\\ 
	& = \sum_{j=1}^{\L-1}H_{j(j+1)} + H_{\L 1}
\end{aligned}
\end{equation}
where $\beta_{1} = \beta_{2}^{*} = \frac{1}{6}\left(3 + i \sqrt{3}\right)$ and
$\alpha_{1}$, $\alpha_{2}$ have to be real for $\mathcal{H}$ to be hermitian.
The local Hamiltonian is
\begin{equation}
H = i \sum_{a,b=1}^{3} \sum_{l=1}^{2} (-1)^{l} \left[ \frac{\alpha_{1}w^{lb} +
    \alpha_{2}w^{-lb}}{\left(w^{-l} - w^{l}\right)}\right] e_{a+b+l,a+b} \tp
e_{a+l,a}, \label{eqnLocalH} 
\end{equation}
with subscripts in (\ref{eqndefglobalperiodicHam}) denoting which spaces the
operator acts upon.  We fix the overall energy scale by setting
$\alpha_{1}=\cos(\theta)$ and $\alpha_{2}=\sin(\theta)$ and separate both the
global and local Hamiltonians into
$$ \mathcal{H} = \cos(\theta) \, \mathcal{H}^{(1)} + \sin(\theta) \,
\mathcal{H}^{(2)} \hspace{0.7cm} \mbox{and} \hspace{0.7cm} H = \cos(\theta) \,
H^{(1)} + \sin(\theta) \, H^{(2)} $$
with the property that 
\begin{eqnarray}
\left[\mathcal{H}^{(1)} ,\,\mathcal{H}^{(2)}\right]=0, 
\label{commute}
\end{eqnarray}
which follows from the transfer matrix $t^{(3,p)}(z_{1},z_{2})$ forming a commuting family in both variables.  
From (\ref{eqnLocalH}) we see that the local Hamiltonians satisfy
$$ \left[ H^{(1)} \right]^{\dagger} = H^{(1)}, \hspace{0.7cm} \left[ H^{(1)} \right]^{*} = H^{(2)} \hspace{0.7cm} \mbox{and} \hspace{0.7cm} \Pi H^{(1)} \Pi = H^{(2)}. $$
The generator of translations is the momentum operator
\begin{equation} \label{eqnMomOp}
	\mathcal{P} = -i\ln T =  -i\ln[t^{(3,p)}(1,1)]
\end{equation}
with eigenvalues being integer multiples of ${2\pi}/{\L}$.  Due to the
periodicity of the model and the non-cocommutativity of the $D(D_3)$ comultiplication
(i.e. ($\Delta\neq \Delta^T$), we find that the global symmetry of the model
is broken due to the $H_{\L 1}$ interaction term, reducing the symmetry to
$D_{3}$ symmetry. The global $D(D_3)$ invariance can be maintained by using a
modified version of the QISM which incorprates generalised versions of the
translation and momentum operators using braiding.
\\

\noindent
\underline{\textbf{Braided closed boundary conditions}}\\
\noindent
To obtain a model with braided closed boundary conditions \cite{lf97} we define 
$$ \bar{L} = L(0) \hspace{0.7cm} \mbox{and} \hspace{0.7cm} \bar{R} = R(0,0). $$
We construct the transfer matrices
$$ t^{(2,b)}(z) = \mbox{tr}_{0}\left[ L_{0\L}(z)...L_{01}(z)\bar{L}_{01}...\bar{L}_{0\L} \right] $$
and
$$ t^{(3,b)}(z_{1},z_{2}) = \mbox{tr}_{0}\left[R_{0\L}(z_{1},z_{2})...R_{01}(z_{1},z_{2})\bar{R}_{01}...\bar{R}_{0\L} \right]. $$
These transfer matrices satisfy the same commutation relations as their
counterparts in the periodic boundary case and have analogue properties. From
$t^{(3,b)}(z_{1},z_{2})$ we construct the global Hamiltonian as was done for the periodic
boundary case, yielding
$$ \mathcal{H} = \sum_{i=1}^{\L -1}H_{i(i+1)} + H_{0}, $$
where $H$ is the local Hamiltonian defined by Equation (\ref{eqnLocalH}) and
$$ H_{0} = GH_{(\L-1)\L}G^{-1}  \hspace{0.7cm} \mbox{with} \hspace{0.7cm} G = t^{(3,b)}(1,1) = \Pi_{21}\bar{R}_{21}...\Pi_{\L (\L-1)}\bar{R}_{\L (\L-1)}\,. $$ 
We have the property that
$$ G H_{i(i+1)}G^{-1} = H_{(i+1)(i+2)}  \hspace{0.7cm} \mbox{and} \hspace{0.7cm} G H_{0} G^{-1} = H_{12} $$
for $i=1...(\L-2)$. For this construction we see that $G$ plays the role of the translation operator, and that $H_{0}$ commutes with all the local Hamiltonians except those which act on the 1st or $\L$th sites. This allows us to define a momentum operator using Equation (\ref{eqnMomOp}).

The operators
$$b_j= \Pi_{(j+1)j}\bar{R}_{(j+1)j}$$
are precisiely the local braiding operators for the anyonic degrees of freedom which satisfy the braid group relations \cite{nssfd2008}.  We can interpret $H_{0}$ as a braided boundary interaction term for sites 1 and $\L$, since the action of $G^{-1}= b^{-1}_{\L-1}...b^{-1}_2b^{-1}_1$ is to braid the state at site 1 through to site $\L$, this state interacts through $H_{(\L-1)\L}$, and then $G=b_1b_2...b_{\L-1}$ braids the state back to site 1.  

Unlike the periodic model case, the global Hamiltonian with braided closed
boundary conditions is invariant under the action of $D(D_{3})$. This is an
important property to maintain in anyonic models as the irreducible
representations of the Hopf symmetry algebra define the global sectors (or
superselection rules) which characterise the total system's ``topological charge'' \cite{kitaev2003,ttwl08}.  
Although the above models differ only in the
boundary conditions, it is not clear that they are equivalent in the
thermodynamic limit since for the braided case the boundary interaction is
non-local.  Our analysis below will be show, however, that the ground states
are equivalent for $\L\to\infty$.   \\

\noindent
\underline{\textbf{Open boundary conditions}}\\
\noindent
We may also preserve the $D(D_3)$ invariance by formulating the model with
open boundary conditions.  In this case the construction of integrable models
can be done within the framework of Sklyanin's reflection algebra
\cite{Sklyanin1988}.  
To this end we require the additional operator 
\begin{equation*}
\begin{aligned}
  \bar{L}(z) &= \sum_{i=1}^{3}\left \{ \left[ e_{1,1} \tp e_{i+1,i} + e_{2,2}
    \tp e_{i-1,i} \right] + z\, \left(w^{i-1}e_{1,2} + w^{1-i}e_{2,1}\right )
  \tp e_{i,i} \right\}\\
  &\propto L^{-1}(-(\omega z)^{-1})\,.
\end{aligned}
\end{equation*}
Within Sklyanin's approach we consider the transfer matrices
\begin{equation*}
\begin{aligned}
 t^{(2,o)}(z) &= \mbox{tr}_{0}\left[
  L_{0\L}(z)...L_{01}(z)\bar{L}_{01}(z)...\bar{L}_{0\L}(z) \right]\,,\\
 t^{(3,o)}(z_{1},z_{2}) &= \mbox{tr}_{0}\left[
  R_{0\L}(z_{1},z_{2})...R_{01}(z_{1},z_{2})R_{10}(z_{1},z_{2})...R_{\L
    0}(z_{1},z_{2}) \right]\,,
\end{aligned}
\end{equation*}
corresponding to specific representations of the reflection algebra based on
unit $K$-matrices describing free ends.
These transfer matrices satisfy the same commutation relations as in the
periodic case and the spectra of $t^{(2,o)}(z)$ and $\left(t^{(2,o)}(z)\right)^{*}$
coincide. By construction $t^{(2,o)}(z)$ will be a polynomial of degree at most $2\L$, similarly the degree of each variable in $t^{(3,o)}(z_{1},z_{2})$ will be also at most $2\L$.

Again a global Hamiltonian can be obtained by Taylor expansion of $t^{(3,o)}(z_{1},z_{2})$.
In this case $t^{(3,o)}(1,1)$ is a scalar multiple of the identity operator, from the first order
term 
we obtain
$$ \mathcal{H} = \sum_{j=1}^{\L-1} H_{j(j+1)}, $$
where $H$ is the local Hamiltonian defined by Eq.~(\ref{eqnLocalH}).
As
with the previous cases both the global and local Hamiltonians can be split
into two coupled Hamiltonians with a coupling parameter $\theta$.   Since
the boundary conditions break translational invariance the notion of a
momentum operator is lost.\\

\noindent
\underline{\textbf{Bethe ansatz solution}}\\
\noindent
%
%
Starting from the Yang--Baxter equation (\ref{eqnYBE}) it is possible to
construct functional relations for the transfer matrices $t^{(2,m)}$ and
$t^{(3,m)}$, $m\in\{p,b,o\}$ \cite{Finch2010}.  In the following the 
fusion relation
\begin{equation} \label{eqnFusion}
 t^{(2,m)}(z_{1})t^{(3,m)}(wz_{1},z_{2}) = f^{(m)}(z_{1})
 t^{(3,m)}(w^{2}z_{1},z_{2}) + g^{(m)}(z_{1})t^{(3,m)}(z_{1},z_{2}). 
\end{equation}
will be used to compute the spectrum of the transfer matrix $t^{(3,m)}$.  Here
the functions $f^{(m)}(z)$ and $g^{(m)}(z)$ are for the periodic, open and
braided models
\begin{eqnarray*}
 f^{(p)}(z) &=& (w^{2}z+1)^{\L}, \\ 
 g^{(p)}(z) &=& (wz-1)^{\L}, \\
 f^{(b)}(z) &=& (1+w^{2}z)^{\L} = f^{(p)}(z), \\
 g^{(b)}(z) &=& (1-wz)^{\L} = (-1)^{\L} g^{(p)}(z), \\
 	f^{(o)}(z) & = & \frac{(1+w^{2}z^{2})(1-wz^{2})}{(1-z^{4})} [1+w^{2}z]^{2\L}, \\
	g^{(o)}(z) & = & (-1)^{\L} \frac{(1-w^{2}z^{2})(1+wz^{2})}{(1-z^{4})}[1-wz]^{2\L}.
\end{eqnarray*}

To obtain the Bethe ansatz solution of the models we proceed by following the
methods of \cite{cdil2010} adapted to the fact that the transfer matrices
$t^{(3,m)}(z_1,z_2)$ are functions of two variables instead of the
single-variable reduction case $z_1=z_2$.  Doing this leads to
the pivotal result


\begin{prop}
Let $\{\lambda^{(m)}_{j}(z) \}$ be the set of eigenvalues of $t^{(2,m)}(z)$ and $\{\phi^{(m)}_{j}(z)\}$ a set of monic polynomials satisfying
\begin{eqnarray}
 \lambda^{(m)}_{j}(z) \phi^{(m)}_{j}(wz) = f^{(m)}(z)\phi^{(m)}_{j}(w^{2}z) + g^{(m)}(z)\phi^{(m)}_{j}(z). 
 \label{fusion} 
 \end{eqnarray}
If a vector $v$ satisfies
$$ t^{(2,m)}(z)v = \lambda^{(m)}_{j}(z)v, \hspace{0.5cm} \left[t^{(2,m)}(z)\right]^{*}v = \lambda^{(m)}_{k}(z)v \hspace{0.5cm} \mbox{and} \hspace{0.5cm} t^{(3,m)}(z_{1},z_{2})v = \Lambda^{(m)}(z_{1},z_{2})v $$
then the general form for $\Lambda^{(m)}(z_{1},z_{2})$ is 
$$ \Lambda^{(m)}(z_{1},z_{2}) = c^{(m)}_{jk} \phi^{(m)}_{j}(z_{1}) \left[\phi_{k}^{(m)}(z_{2})\right]^*, $$
for $z_1,z_2\in{\mathbb{R}}$ where $c^{(m)}_{jk}$ is some constant (not uniquely defined by $j$ and $k$).
\end{prop}

\noindent
It should be noted that the functions $\phi_{j}^{(m)}(z)$ are not uniquely defined. However, if we restrict the degree of the $\phi_{j}^{(m)}(z)$ based upon the degree of $t^{(3,m)}(z_{1},z_{2})$ then the $\phi_{j}^{m}(z)$ have been observed to be unique.

For ease of notation, we hereafter drop the superscript $(m)$ on the
understanding that all subsequent relations hold for a fixed value of  $m \in
\{p,b,o\}$ unless noted otherwise. 
From the above Propositon we see that the energy of the global Hamiltonian for the periodic and braided closed models will be of the form
\begin{eqnarray*}
	E & = & \alpha_{1} i \left[ \phi_{j}^{-1}(1)\phi_{j}^{'}(1) -
          \frac{1}{6}\left(3 + i \sqrt{3}\right) \L \right] - \alpha_{2} i
        \left[ \phi_{k}^{-1}(1)\phi_{k}^{'}(1) - \frac{1}{6}\left(3 + i
            \sqrt{3}\right) \L \right]^{*}. 
\end{eqnarray*}
Noting that all of the global Hamiltonians are self-adjoint and that the energies of each will be of a similar form, we find that energies can be expressed as
%
\begin{equation}
\label{eqnEij}
\begin{aligned}
  E &= \alpha_{1} E_{j} + \alpha_{2} E_{k}\\
 & \mbox{where} \quad
  E_{j}^{(m)} = \left\{
   \begin{array}{ll}
     i \left[ \phi_{j}^{-1}(1)\phi_{j}^{'}(1) - 
  \frac{1}{6}\left(3 + i \sqrt{3}\right) \L \right] 
  & \mbox{for~}m=p,b\,,\\[10pt] 
     i \left[ \frac{1}{2}\,\phi_{j}^{-1}(1)\phi_{j}^{'}(1) - 
  \frac{1}{6}\left(3 + i \sqrt{3}\right) \L \right] 
  & \mbox{for~} m=o\,.
   \end{array}
  \right.
\end{aligned}
\end{equation}
Also of interest to us is the corresponding momentum of an eigenstate in the
case of the periodic and braided closed boundary models. The momentum is given
by 
\begin{equation}
 P \equiv \left(P_{j} - P_{k}^{*} -i\ln(c_{jk})\right) \mod 2 \pi \hspace{0.7cm} \mbox{where} \hspace{0.7cm} P_{j} = -i \ln(\phi_{j}(1)). 
\label{mom}
\end{equation}
We remark that while $P$ must be real it is not guaranteed that the individual
$P_{j}$ will be real. 

To proceed with the solution of the eigenvalue problem we have to address two
problems:  
First, we need to determine the polynomials $\phi_{j}(z)$ occuring in the
solution given in Proposition 3.1.  Rewriting (\ref{fusion}) as
\begin{eqnarray*}
 \lambda_{j}(z)  = f(z)\frac{\phi_{j}(w^{2}z)}{\phi_{j}(wz)} + g(z)\frac{\phi_{j}(z)}{\phi_{j}(wz)}
\end{eqnarray*}
and using the fact that this is an equality of polynomials, the residues
evaluated at the zeroes of $\phi_{j}(wz)$ in the right hand expression must
vanish.  Therefore, characterizing $\phi_j(z)$ by the set of its zeroes $z_k$
we obtain the Bethe  equations for each model as
\begin{eqnarray}
 \lim_{z\rightarrow z_k}(z-z_k)\lambda_{j}(\omega^2 z) &=& 0
 \nonumber\\
 \Rightarrow \frac{f(\omega^2 z_k)}{g(\omega^2
   z_k)}&=&-\frac{\phi(\omega^2z_k)}{\phi(\omega z_k)}.
 \label{eqnBAE}
\end{eqnarray}
Each set of non-degenerate roots $\{z_k\}$ of these equations provides us with
a polynomial $\phi_{j}(z)$.

Second, we need to determine a \textit{pairing} rule that states whether the
polynomials $\phi_{j}(z_1)$ and $\phi_{k}(z_2)$ can be combined to eigenvalues
$\Lambda(z_1,z_2)$ of the transfer matrix $t^{(3)}(z_1,z_2)$  according to
Proposition 3.1.  Only pairs $(i,j)$ which are allowed by this rule will
determine energy and momentum of an eigenstate of the global Hamiltonian.


\section{Observations from numerical results of models with few sites}
\label{secFewSiteModels}
In this section we begin to study these questions numerically for models with
a few lattice sites.  In these systems we can identify the root configurations
to the Bethe equations (\ref{eqnBAE}) corresponding to the ground states of
the system and, even more important, the patterns shown by these
configurations which allow for an analysis of the spectrum in the
thermodynamic limit $\L\to\infty$.  Furthermore, our finite lattice analysis
provides us with a basis to conjecture the pairing rules mentioned above.


\noindent
For each model we consider the following ansatz
$$ \phi_{j}(z) = \prod_{k=1}^{d_{j}}(z - iwy_{jk}), $$
for some non-zero $d_{j} \in \N$, or $\phi_{j}(z)=1$ for $d_j=0$.. 
According to (\ref{eqnEij}) the energy associated with this
function is given by
\begin{equation*}
 E_{j}^{(m)} = \left\{
   \begin{array}{ll}
   i\sum_{k=1}^{d_{j}} \frac{1}{1-iwy_{jk}} - \frac{i}{6}\left(3 +
   i\sqrt{3} \right)\L & \mbox{for~}m=p,b\,,\\[10pt]
   \frac{i}{2}\sum_{k=1}^{d_{j}} \frac{1}{1-iwy_{jk}} - \frac{i}{6}\left(3 +
   i\sqrt{3} \right)\L & \mbox{for~}m=o\,. 
\end{array}\right.
\end{equation*}
Similarly, from (\ref{mom}) we can express the momentum for the closed
boundary conditions in terms of the Bethe
roots,
$$ P_{j} = -i \sum_{k=1}^{d_{j}} \ln \left( 1-iwy_{jk} \right). $$

We comment here that the Bethe ansatz solutions below are incomplete in the
sense that they determine the eigenvalues of the transfer matrix
$t^{(3)}(z_1,z_2)$ only up to the factor $c_{jk}$, as they only give constraints
on the variables $y_{jk}$.  This does not restrict its use for studies of the
spectrum of the Hamiltonian since the energy expression $E_{j}$ is not
dependent on the $c_{jk}$. The total momentum (\ref{mom}), however, does
depend on $c_{jk}$. We have found through our numerical studies that the
$c_{jk}$ are always real (they may be either positive or negative). Therefore
their contribution to (\ref{mom}) is zero modulo $\pi$, which is sufficient to
determine whether a particular state is time-reversal invariant
(i.e. invariant under complex conjugation, up to a phase) or not.

The Bethe  equations that we present also admit spurious solutions \cite{cdil2010},
i.e. there are solutions  which do not correspond to actual eigenvalues of the
transfer matrices. Below, the statements concerning the properties of the
Bethe roots relate to solutions which are not spurious solutions. \\ 

\noindent
\underline{\textbf{Periodic boundary conditions}}\\
\noindent
Using our ansatz for $\phi_{j}(z)$  the Bethe equations become
$$ \prod_{k=1}^{d_{j}} \left(\frac{wy_{jl}-y_{jk}}{w^{2}y_{jl}-y_{jk}}\right) = -\left(\frac{iwy_{jl}-1}{iw^{2}y_{jl}+1}\right)^{\L}, $$
for $0 \leq l \leq d_{j}$.
These are the same equations found in \cite{cdil2010}.
\begin{obs}
We make the following observations about the Bethe roots of the function $\phi_{j}(z)$,
\begin{enumerate}
	\item if $y_{jk}$ is a Bethe root then so is $y_{jk}^{*}$,
	\item there are $\L$ or $\L-1$ distinct Bethe roots, one of which may be zero.
	\item any two functions $\phi_{j}(z)$ and $\phi_{k}(z)$ may pair if and only if both functions have the same number of non-zero and zero roots.
\end{enumerate}
\end{obs}

\noindent
Of particular importance are the possible ground states, i.e.\ the eigenstates
of $\mathcal{H}^{(1)}$ (or equivalently $\mathcal{H}^{(2)}$) with highest and
lowest energy.  Our numerical results for the corresponding configurations of
Bethe roots for the highest energy state in a system with $\mathcal{L}=2,3,4$
are shown in Table~\ref{tblperhigh}.
%
\begin{table}[ht] 

\caption{\label{tblperhigh}
Bethe roots for highest energy state of $\mathcal{H}^{(1)}$}

$$ \begin{array}{|c|c|c|c|} 
\hline
\L & \mbox{Energy} & y_{jk} & \ln(y_{jk}) \\
\hline 
2 & 2.31 & -0.57735+0.81650 & 2.1863i \\
&& -0.57735-0.81650 & -2.1863i \\
\hline 
3 & 1.97 & -0.38896+0.76875i & -0.14902+2.0392i \\
&& -0.38896-0.76875i & -0.14902-2.0392i \\
\hline
4 & 3.15 & -0.74151+1.1157i & 0.29240+2.1574i \\
&& -0.74151-1.1157i & 0.29240-2.1574i \\
&& -0.41319+0.62170i & -0.29238+2.1574i \\
&& -0.41319-0.62170i & -0.29238-2.1574i \\ \hline
\end{array} $$

\end{table} 
\noindent
From these data we conjecture that this state is determined by complex
conjugate pairs of roots, the $\ln(y_{jk})$ form so-called 2-strings \cite{Bethe31,Taka71b}.  The
pairs are centered on the real line and seperated by ${4\pi i}/{3}$ up to
finite-size effects.  Our data for $\L=5$ and $6$ support this conjecture and
we have verified numerically that such a configuration solves the Bethe
equations for lattices with up to $\L\approx1000$ sites.\\

The Bethe roots for the lowest energy states are given in
Table~\ref{tblperlow}. In this table we have included a case where two
different sets of Bethe roots give the same energy. \\

\begin{table}[ht] 
\caption{\label{tblperlow}
Bethe roots for lowest energy state of $\mathcal{H}^{(1)}$} 
$$ \begin{array}{|c|c|c|c|} 
\hline
\L & \mbox{Energy} & y_{jk} & \ln(y_{jk}) \\
\hline 
2 & -1.15 & -0.86603 & -0.14384 + i\pi \\
\hline 
2 & -1.15 & 0 & -\infty \\
&& -1.1547 & 0.14384 + i\pi \\
\hline 
3 & -1.73 & 0.50771 & -0.67784 \\
&& -0.77786 & -0.25121 + i\pi \\
&& -1.4619 & 0.37974 + i\pi \\
\hline
4 & -2.31 & 1 & 0 \\
&& -1 & i\pi \\
&& -1.7321 &0.54933+i\pi \\
&& -0.57735 &-0.54933+i\pi \\ \hline
\end{array} $$
\end{table} 

\noindent
We observe that all of the Bethe roots $y_{jk}$ are real, or equivalently, the
logarithm of all Bethe roots lie either on the real line or are shifted by
$i\pi$.  In the case of $\L \equiv 0 \mod 4$ it has been found that the set of
Bethe roots is invariant under inversion $\{y_{jk}\}\leftrightarrow
\{1/y_{jk}\}$ with $\L/4$ positive and $3\L/4$  negative Bethe roots. 
%
%
Again we have obtained data for $\L$ up to 640 sites in support of this
observation.  \\


\noindent
\underline{\textbf{Braided closed boundary conditions}}\\
\noindent
Using our ansatz for $\phi_{j}(z$) the Bethe equations become
$$ \prod_{k=1}^{d_{j}} \left( \frac{wy_{jl}-y_{jk}}{w^{2}y_{jl}-y_{jk}}
\right) = - \left(\frac{1-iwy_{jl}}{iw^{2}y_{jl}+1}\right)^{\L}. $$ 
For $\L$ even these coincide with the equations for the periodic case,
although the conditions imposed on non-spurious solutions are different:  
\begin{obs}
We make the following observations about the Bethe roots of the function
$\phi_{j}(z)$, 
\begin{enumerate}
	\item if $y_{jk}$ is a Bethe root then so is $y_{jk}^{*}$,
	\item there are $\L$ or $\L-1$ distinct Bethe roots all of which are
          non-zero, 
	\item if $\lambda(z)$ is an eigenvalue of $t^{(2,b)}(z)$ then
          $z^{\L}\lambda(z^{-1})$ is an eigenvalue of $t^{(2,p)}(z)$, 
	\item any two functions $\phi_{j}(z)$ and $\phi_{k}(z)$ may pair, 
	\item the energy spectrum $\mathcal{H}^{(1)}$ with braided closed
          boundary conditions is a subset of the energy spectrum of
          $\mathcal{H}^{(1)}$  with periodic boundary conditions. 
\end{enumerate}
\end{obs}
\noindent
Note that while the third observation implies the fifth one it does not
provide any insights into the degeneracies of the spectrum for the global
Hamiltonian $\mathcal{H}$.
The difference in pairing rules compared to the periodic boundary model
is related to the change in the symmetry of the global Hamiltonian
i.e.\ the difference between $D_{3}$ and $D(D_{3})$ symmetry.
The condition that all Bethe roots be non-zero renders some of the
configurations found in the case of periodic boundaries spurious.  The
solutions to the Bethe equations with $y_{jk}\ne0$ for all $k$ for the ground
states identified above, however, do correspond to ground states in the
presence of braided closed boundary conditions. \\




\noindent
\underline{\textbf{Open boundary conditions}}\\
\noindent
Using the ansatz for $\phi_{j}(z)$ the Bethe equations become
$$ \prod_{k=1}^{d_{j}}\left(\frac{wy_{jl}-y_{jk}}{w^{2}y_{jl}-y_{jk}}\right) =
(-1)^{\L+1}\left(\frac{1-wy_{jl}^{2}} {1-w^{2}y_{jl}^{2}}\right)
\left(\frac{1+w^{2}y_{jl}^{2}} {1+wy_{jl}^{2}}\right) \left(\frac{1-iwy_{jl}}
  {1+iw^{2}y_{jl}}\right)^{2\L}, $$ 
for $0 \leq l \leq d_{j}$.
\begin{obs}
We make the following observations about the Bethe roots of the function $\phi_{j}(z)$,
\begin{enumerate}
	\item if $y_{jk}$ is a Bethe root then so is $y_{jk}^{*}$,
	\item if $y_{jk}$ is a non-zero Bethe root then so is $y_{jk}^{-1}$,
	\item there are $2\L$ or $2\L-1$ distinct Bethe roots,
	\item if there are $2\L$ Bethe roots they are all non-zero,
	\item if there are $2\L-1$ Bethe roots one of them is zero,
	\item any two functions $\phi_{j}(z)$ and $\phi_{k}(z)$ may pair.
\end{enumerate}
\end{obs}
\noindent
Our numerical analysis of the Bethe equations for systems with a few sites
shows that apart from the doubled number of roots the ground states are again
described by 2-strings and real $y_{jk}$, repsectively.  The existence of
these solutions has been verified for systems with several hundred sites.

\section{Ground-state energy in the thermodynamic limit}
We now calculate the ground state energy in the thermodynamic limit for the
periodic, braided closed boundary, and open models.  We begin by considering
the states with highest and lowest energies of the Hamiltonian
$\mathcal{H}^{(1)}$.  From our analysis of small systems we know that the
Bethe root configurations for the highest and lowest energy states for the
different boundary conditions are essentially the same (apart from the
doubling of roots in the open boundary case).

Expanding the energy in $\L$ one has the general form
$$ E = \L \epsilon_{\infty} + \rho_{\infty} + \frac{1}{\L} \times
\mbox{const.} + {o}(\L^{-1}), $$ 
where $\L \epsilon_{\infty}$ and $\rho_{\infty}$ are the bulk and boundary
energy contributions, respectively.\\

\noindent
\underline{\textbf{Bulk Energy of $\mathcal{H}^{(1)}$}}\\
\noindent
Here we calculate the bulk energy for the highest and lowest energy states of
$\mathcal{H}^{(1)}$ in the thermodynamic limit $\L\to\infty$.  By definition
this term is linear in the system size $\L$ and independent of boundary
contributions.  Below we assume $\L$ to be even, therefore our results apply
to chains with both periodic and braided closed boundary conditions having the
same Bethe equations.\\ 

\noindent
Motivated by our numerical results for the highest energy state of the model
with a small even number of lattice sites, we consider a distribution
of Bethe roots given by
$$ y_{2k-1} = e^{x_{k}+\frac{2i\pi}{3}} \hspace{0.7cm} \mbox{and}
\hspace{0.7cm} y_{2k} = e^{x_{k}-\frac{2i\pi}{3}} $$ for $1 \leq k \leq
{\L}/{2}$ with $x_{k}\in \R$.  This configuration of Bethe roots is the
aforementioned 2-string hypothesis.  In terms of the new variables $x_{k}$ the
Bethe equations become
$$ \prod_{k=1}^{\frac{\L}{2}}\left[\frac{\sinh(\frac{x_{l}
      -x_{k}}{2}+\frac{i\pi}{3})}{\sinh(\frac{x_{l}
      -x_{k}}{2}-\frac{i\pi}{3})} \right] 
= \left[  \frac{\sinh(\frac{x_{l}}{2} +
    \frac{15i\pi}{12})\sinh(\frac{x_{l}}{2} +
    \frac{11i\pi}{12})}{\sinh(\frac{x_{l}}{2} -
    \frac{11i\pi}{12})\sinh(\frac{x_{l}}{2} - \frac{15i\pi}{12})}
\right]^{\L}, $$ 
for $0 \leq l \leq {\L}/{2}$. 
%
%
For $\L\to\infty$ the roots $x_k$ can be described by their density
$\rho(x)$ which, after a Fourier transformation, satisfies the linear
equation
$$ \tilde{\rho}(v) = - \tilde{B}\left(v;\frac{i\pi}{4}\right) - \tilde{B}\left(v;\frac{11i\pi}{12}\right) + \tilde{B}\left(v;\frac{i\pi}{3}\right)\tilde{\rho}(v), $$
where
 $$ \tilde{B}(v,t) = \frac{-1}{2i\pi}\int_{-\infty}^{\infty}\left[\frac{d}{dx} \ln \left( \frac{\sinh(\frac{x}{2}+t)}{\sinh(\frac{x}{2}-t)} \right)\right]e^{-ivx}dx = \frac{\sinh(v(\pi+2it))}{\sinh(\pi v)}. $$
Solving this we find the density
\begin{eqnarray*}
	\rho(x)	& = & \frac{3}{2\pi} \frac{1}{\cosh(3x)}.
\end{eqnarray*}
We calculate the energy to be
\begin{eqnarray*}
	\L \epsilon_{\infty} 
	& = & i\sum_{k=1}^{\frac{\L}{2}} \left[\frac{1}{1-ie^{x_{k}}} + \frac{1}{1-ie^{x_{k}+\frac{4i\pi}{3}}}\right] - \frac{i}{6}\L(3+i\sqrt{3}) \\
	& = & i \L \int_{-\infty}^{\infty} \rho(x)\left[\frac{1}{1-ie^{x}} + \frac{1}{1-ie^{x+\frac{4i\pi}{3}}}\right] dx - \frac{i}{6}\L(3+i\sqrt{3}) \\
	& = &  \L \left[ \frac{1}{\pi} + \frac{2\sqrt{3}}{9} \right]\,.
\end{eqnarray*}
This is the bulk energy of the highest energy state of $\mathcal{H}^{(1)}$. \\

\noindent
For the lowest energy state of $\mathcal{H}^{(1)}$ we consider chains of
length $\L\equiv0\mod 4$ where according to our conjecture above the root
configuration is given by
\begin{eqnarray*}
	y_{k}=e^{x_{k}^{a}}, &\hspace{0.3cm}& 1 \leq k \leq {\L}/{4}, \\
	y_{k+\frac{\L}{4}}=e^{x_{k}^{b}+i\pi}, && 1 \leq k \leq {3\L}/{4},
\end{eqnarray*}
with $x_{k}^{a},x_{k}^{b} \in \R$.  We define two different density functions
$\rho_{a,b}$ for the two subsets of Bethe roots.  From the Bethe equations we
can construct coupled equations which are linear in Fourier space.  Solving
these equations yields
\begin{eqnarray*}
	\rho_{a}(x)
	& = & \frac{3}{4} \frac{\sqrt{2}\cosh (\frac{3}{2}x)}{\pi\cosh(3x)} - \frac{3}{4}\frac{1}{\pi\cosh(3x)}
\end{eqnarray*}
and
\begin{eqnarray*}
	\rho_{b}(x)
	& = & \frac{3}{4} \frac{\sqrt{2}\cosh (\frac{3}{2}x)}{\pi\cosh(3x)} + \frac{3}{4}\frac{1}{\pi\cosh(3x)}.
\end{eqnarray*}
With these densities we obtain for the bulk energy of the lowest energy state
of $\mathcal{H}^{(1)}$
\begin{eqnarray*}
	\L \epsilon_{\infty}
	& = & i \L \left[\int_{-\infty}^{\infty} \left(\frac{\rho_{a}(x)}{1-iwe^{x}} + \frac{\rho_{b}(x)}{1+iwe^{x}}\right)dx - \frac{1}{6}\left(3 + i\sqrt{3} \right)\right] \\
	& = & -\L\left(\frac{1}{2\pi} - \frac{2\sqrt{3}}{9} + \frac{3}{4}\right)\,.
\end{eqnarray*}

\noindent
\underline{\textbf{Boundary Energy of $\mathcal{H}^{(1)}$}}\\
\noindent
Next we compute the boundary energy $\rho_\infty$ for these states.  As a
consequence of translational invariance $\rho_\infty$ vanishes in the closed
chains.  We can deal with open boundaries as above with two modifications: due
to Observation 4.3 we now have to consider $2\L$ roots which for the highest
energy state of $\mathcal{H}^{(1)}$ amounts to parametrizing the roots as
$$ y_{2k-1} = e^{x_{k}+\frac{2i\pi}{3}} \hspace{0.7cm} \mbox{and}
\hspace{0.7cm} y_{2k} = e^{x_{k}-\frac{2i\pi}{3}} $$ for $1 \leq k \leq \L$
with $x_{k}\in \R$.  In addition only even distributions of the $x_k$ are to
be considered (see also \cite{AsSu96,HaQB87}).  Proceeding as above we
introduce a density function $\rho(x)$ which in Fourier space satisfies
\begin{eqnarray*}
	\tilde{\rho}(v)  &=&
         - 2\left(\tilde{B}\left(v;\frac{i\pi}{4}\right) +
          \tilde{B}\left(v;\frac{11i\pi}{12}\right)\right) \\
	& &
        -\frac{1}{\L}\left(\tilde{B}\left(\frac{v}{2};\frac{2i\pi}{3}\right) +
          \tilde{B}\left(\frac{v}{2};\frac{5i\pi}{6}\right) +
          \tilde{B}\left(v;\frac{i\pi}{3}\right)\right)+
        \tilde{B}\left(v;\frac{i\pi}{3}\right)\tilde{\rho}(v).  
\end{eqnarray*}
To order $\L^0$ this equation is solved by twice the density function found
for the closed boundary conditions above.  This reproduces the bulk energy of
the highest energy state.  The order $\L^{-1}$ correction of the density gives
the boundary energy for this state
$$ \rho_{\infty} = \left[ -\frac{3}{2} + \frac{2\sqrt{3}}{3} \right]. $$
\\

\noindent
For the calculation of the boundary contribution to the lowest energy state of
$\mathcal{H}^{(1)}$ we proceed in the
same way.  As a result we obtain

$$ \rho_{\infty} = \left[-\frac{3}{4} + \frac{2\sqrt{3}}{3} \right]. $$ \\
\\

\noindent
\underline{\textbf{Energy, momentum and degeneracy of the ground state}}\\
For the highest and lowest energy states we have calculated both the bulk and
boundary energy analytically for $\mathcal{H}^{(1)}$.  Our numerical studies
of small systems indicate that the excitation spectrum is gapless.
The finite size corrections (of order $\L^{-1}$) to the ground state energies
can be obtained through numerical analysis of the Bethe equations for systems
with up to $\L=1000$ lattice sites.
%
%
For the closed models we find that the highest energy is
$$ E_{\mbox{high}} = \left[ \frac{1}{\pi} + \frac{2\sqrt{3}}{9} \right] \L +  \frac{12}{5}\times \frac{\pi}{6\L} + {o}(\L^{-1}), $$
while the lowest energy is
$$ E_{\mbox{low}} = -\left[\frac{1}{2\pi} - \frac{2\sqrt{3}}{9} + \frac{3}{4}\right] \L - \frac{3}{2}\times \frac{\pi}{6\L} + {o}(\L^{-1}). $$
Similarly, we present the energies for the open model. The highest energy is
$$ E_{\mbox{high}} = \left[ \frac{1}{\pi} + \frac{2\sqrt{3}}{9} \right] \L + \left[ -\frac{3}{2} + \frac{2\sqrt{3}}{3} \right] + \frac{12}{5}\times \frac{\pi}{24\L} + {o}(\L^{-1}), $$
while the lowest energy is
$$ E_{\mbox{low}} = -\left[\frac{1}{2\pi} - \frac{2\sqrt{3}}{9} + \frac{3}{4}\right] \L + \left[-\frac{3}{4} + \frac{2\sqrt{3}}{3} \right] - \frac{3}{2}\times \frac{\pi}{24\L} + {o}(\L^{-1}). $$
Note that this finite size scaling behaviour is consistent with the
predictions of conformal field theory \cite{Affl86,BlCN86}.  To identify the
Virasoro algebras corresponding to these sectors of the model we need to
include the low-energy excitations in our analysis.  This will allow us to
compute the (non-universal) Fermi-velocity needed to extract the central
charge from the $\L^{-1}$-corrections to the ground state energies and also
the conformal dimensions appearing in the excitation spectrum.  These
questions will be addressed in future work.

From the previously stated properties we know that the energies given above
must also be the highest and lowest energies of $\mathcal{H}^{(2)}$.
Furthermore, based on our analysis of the models with few sites we expect that
both the highest and lowest energies pair.  Therefore, the ground-state energy
will depend upon the values of the coupling parameter $\theta$ and is
specifically given by,
$$ E_{\mbox{ground state}} 
= \left\{ \begin{array}{lrcl} 
\cos(\theta)E_{\mbox{low}} + \sin(\theta)E_{\mbox{low}}, & 0 \leq & \theta & < \frac{\pi}{2}, \\
\cos(\theta)E_{\mbox{high}} + \sin(\theta)E_{\mbox{low}}, & \frac{\pi}{2} \leq & \theta & < \pi, \\
\cos(\theta)E_{\mbox{high}} + \sin(\theta)E_{\mbox{high}}, & \pi \leq & \theta & < \frac{3\pi}{2}, \\
\cos(\theta)E_{\mbox{low}} + \sin(\theta)E_{\mbox{high}}, & \frac{3\pi}{2} \leq & \theta & < 2\pi
\end{array} \right. $$
where we have made use of (\ref{commute}). 

Due to the manner in which all the $\phi_k(z)$ pair to produce the transfer matrix spectrum when the $D(D_3)$ symmetry is not 
broken, it is anticipated that the number of (non-spurious) solutions of the Bethe ansatz equations scales exponentially with $\L$. Our numerical calculations support this view. It is clear that level crossings will occur when $\theta = m{\pi}/{2}$ for $m \in \Z$ yielding either $\alpha_1=0$ or $\alpha_2=0$. These level crossings  at which the ground state is degenerate give first order quantum phase transitions, as it is straightforward to show that the energy has discontinuous first derivative. The ground-state energy at these transition points is determined by a single solution of the Bethe ansatz equations, with the second solution in (\ref{eqnEij}) chosen arbitrarily. Hence the ground-state degeneracy will depend on the number of solutions of the Bethe ansatz equations, leading to an unusual exponential scaling of the degeneracy. 

While the ground state is non-degenerate for $\theta \neq m{\pi}/{2}$, it is important to briefly discuss the fact that $D(D_3)$ admits two distinct one-dimensional representations \cite{DIL2006} which define two different global topological charge sectors. While we cannot determine which of these sectors the ground state is a member of generally, we can establish that the ground state for ${\pi}/{2} <  \theta  < \pi$ and the ground state for ${3\pi}/{2} <  \theta  < 2\pi$ belong to the same sector. This follows from the properties i) $[\mathcal{H}^{(1)}]^*=\mathcal{H}^{(2)}$, meaning that the ground state for ${\pi}/{2} <  \theta  < \pi$ is the complex conjugate of the ground state for ${3\pi}/{2} <  \theta  < 2\pi$,  and ii)  complex conjugation leaves all topological charge sectors invariant since the representation (\ref{rep}) is real. To show that the ground states for ${\pi}/{2} <  \theta  < \pi$ and ${3\pi}/{2} <  \theta  < 2\pi$ are distinct in the case of the closed models, it remains to finally consider the momentum.

Using the densities of Bethe roots calculated above  we can compute the momentum of the ground state for the closed chain models:
$$ \mbox{Re}(P_{\mbox{low}}) = -\frac{7\pi}{24} \L \hspace{0.7cm} \mbox{and} \hspace{0.7cm}  \mbox{Re}(P_{\mbox{high}}) = \frac{\pi}{12} \L. $$
We have ignored the imaginary components as they have been observed to be cancelled out by the imaginary component contributed by $c_{jk}$. Furthurmore we  observed that the constants $c_{jk}$ are real, this determines the momentum modulo $\pi$:
$$ P_{\mbox{ground state}} 
\equiv \left\{ \begin{array}{clrcl} 
0 & \mod \pi, & 0 < & \theta & < \frac{\pi}{2}, \\
\frac{3\pi}{8} \L & \mod \pi, & \frac{\pi}{2} < & \theta & < \pi, \\
0 & \mod \pi, & \pi < & \theta & < \frac{3\pi}{2}, \\
-\frac{3\pi}{8} \L & \mod \pi, & \frac{3\pi}{2} < & \theta & < 2\pi.
\end{array} \right. $$ \\
From the above we see that a signature of the ground-state phases for ${\pi}/{2} <  \theta  < \pi$ and ${3\pi}/{2} <  \theta  < 2\pi$ is that the momentum is generally non-zero modulo $\pi$, so the states are not time-reversal invariant. Thus these two phases are two distinct chiral phases that belong to the same topological charge sector. Conversely the ground states for $0 <  \theta  < \pi/2$ and $ \pi <  \theta < {3\pi}/{2} $ are time-reversal invariant.


\section{Summary}

The Hamiltonian we have studied admits gapless excitations in the thermodynamic limit for all values of the coupling parameter $\theta$. At the particular values $\theta = m{\pi}/{2}$ for $m \in \Z$ ground-state energy level crossings occur, which divide the ground-state phase diagram into four regions. We have computed the ground-state energy for each of these regions for both closed and open boundary conditions. From these explicit expressions it is seen that the level crossing points correspond to first order transitions where the first derivative of the ground-state energy is discontinuous. We also computed the ground-state momentum in the closed chain case, to establish that chiral phases exist which are characterised by non-vanishing momentum.


\begin{thebibliography}{99}

\bibitem{Affl86}
I. Affleck, \textit{Universal term in the free energy at a critical point
  and the conformal anomaly}, Phys. Rev. Lett. \textbf{56}, 746--748 (1986).

\bibitem{AsSu96}
H. Asakawa and M. Suzuki, \textit{Finite-size corrections in the {XXZ} model
  and the {H}ubbard model with boundary fields}, J. Phys. A: Math. Gen. \textbf{29},
  225--245 (1996).
  
\bibitem{bpa88} 
R.J. Baxter, J.H.H. Perk, and H. Au-Yang, \textit{New solutions of the star-triangle relations for the chiral Potts model}, Phys. Lett. A \textbf{128}, 138--142, (1988).

\bibitem{Bethe31}
H. Bethe, \textit{{Zur {T}heorie der {M}etalle. I. {E}igenwerte und
  {E}igenfunktionen der linearen {A}tomkette}}, Z. Phys. \textbf{71}, 205--226
  (1931).

\bibitem{BlCN86}
H.W.J. Bl{\"o}te, J.L. Cardy and M.P. Nightingale, \textit{Conformal
  invariance, the central charge and universal finite-size amplitudes at
  criticality}, Phys. Rev. Lett. \textbf{56}, 742--745 (1986).

\bibitem{by07} 
N.E. Bonesteel and K. Yang, \textit{Infinite-randomness fixed points for chains of non-Abelian quasiparticles}, Phys. Rev. Lett. \textbf{99}, 140405 (2007). 

\bibitem{bjjst06} 
H. Boos, M. Jimbo, T. Miwa, F. Smirnov, and Y. Takeyama, 
\textit{Algebraic representation of correlation functions in integrable spin chains},
Ann. Henri Poincare \textbf{7}, 1395--1428 (2006).    


\bibitem{cdil2010}
C.W. Campbell, K.A. Dancer, P.S. Isaac and J. Links, \textit{Bethe ansatz
  solution of an integrable, non-Abelian anyon chain with $D(D_3)$ symmetry},
  Nucl. Phys. B \textbf{836}, 171�185, (2010). 

\bibitem{DIL2006}
K.A. Dancer, P.S. Isaac and J. Links, \textit{Representations of the quantum
  double of finite group algebras and spectral parameter dependent solutions of
  the Yang--Baxter equation}, J. Math. Phys., \textbf{47}, 103511, (2006).

\bibitem{d1986} 
V.G. Drinfeld,  \textit{Quantum groups} in Proceedings of the International Congress of
Mathematicians, A.M. Gleason (ed.) pp. 798-820 (Providence, RI: American Mathematical
Society, 1986). 

\bibitem{fz82} V. Fateev and A.B. Zamolodchikov, \textit{Self-dual solutions of the star-triangle relations in $\mathbb{Z}_N$ models}, Phys. Lett. A \textbf{92}, 37--39, (1982). 
     
\bibitem{ftltkwf2007}
A. Feiguin, S. Trebst, A.W.W. Ludwig, M. Troyer, A. Kitaev, Z. Wang, and M.H. Freedman,
\textit{Interacting anyons in topological quantum liquids: the golden chain}, Phys. Rev. Lett. \textbf{98}, 160409, (2007). 

\bibitem{frbm08}  
L. Fidkowski, G. Refael, N.E. Bonesteel, and J.E. Moore,
{\it c-theorem violation for effective central charge of infinite-randomness fixed points},
Phys. Rev. B \textbf{78}, 224204, (2008).  

\bibitem{Finch2010}
P.E. Finch, \textit{Integrable Hamiltonians with $D(D_n)$ symmetry from the Fateev-Zamolodchikov model}, in preparation, (2010).

\bibitem{FDIL2010}
P.E. Finch, K.A. Dancer, P.S. Isaac and J. Links, \textit{Solutions of the
  Yang-Baxter equation: descendants of the six-vertex model from the Drinfeld
  doubles of dihedral group algebras}, unpublished (arXiv:1003.0501), (2010).

\bibitem{gatltw2009} 
C. Gils, E. Ardonne, S. Trebst, A.W.W Ludwig, M. Troyer and Z.H. Wang, 
{\it  Collective states of interacting anyons, edge states, and the nucleation of topological liquids}, Phys. Rev. Lett. {\bf 103},  070401 (2009). 

\bibitem{HaQB87}
C.J. Hamer, G.R.W. Quispel and M.T. Batchelor, \textit{Conformal anomaly and
  surface energy for Potts and Ashkin-Teller quantum chains}, J. Phys. A: Math. Gen.,
  \textbf{20}, 5677-5693, (1987).

\bibitem{kitaev2003}
A.Y. Kitaev, \textit{Fault-tolerant quantum computation by anyons}, Ann. Phys.,
  \textbf{303}, 2-30, (2003).


\bibitem{kkmst09} 
N. Kitanine, K.K. Kozlowski, J.M. Maillet, N.A. Slavnov, and V. Terras,
\textit{Algebraic Bethe ansatz approach to the asymptotic behavior of correlation functions},
J. Stat. Mech.: Theor. Exp. P04003 (2009).
    

 
 \bibitem{k04}
A. Kl\"umper,
\textit{Integrability of quantum chains: theory and applications to the spin-1/2 $XXZ$ chain} 
Lect. Notes Phys. \textbf{645}, 349--379, (2004).

\bibitem{VladB} 
V.E. Korepin, N.M. Bogoliubov and A.G. Izergin, \textit{Quantum inverse
  scattering method and correlation functions}, Cambridge University Press,
  (1993).

\bibitem{lf97}
J. Links and A. Foerster, 
\textit{On the construction of integrable closed chains with quantum supersymmetry}, 
J. Phys. A: Math. Gen. \textbf{30}, 2483-2487, (1997).
  
\bibitem{mc10}
 J. Mossel and J.-S. Caux,
 \textit{Relaxation dynamics in the gapped $XXZ$ spin-1/2 chain}, 
New J. Phys. \textbf{12}, 055028, (2010).
   

\bibitem{nssfd2008}
C. Nayak, S.H. Simon, A. Stern, M. Freedman, and S. Das Sarma, 
\textit{Non-Abelian anyons and topological quantum computation},  
Rev. Mod. Phys. \textbf{80}, 1083--1159, (2008).

\bibitem{pschmwa06} 
R.G. Pereira, J. Sirker, J.-S. Caux, R. Hagemans, J.M. Maillet, S.R. White, and I. Affleck,
\textit{The dynamical spin structure factor for the anisotropic spin-1/2 Heisenberg chain}, 
Phys. Rev. Lett. \textbf{96}, 257202, (2006). 

\bibitem{Sklyanin1988}
E.K. Sklyanin, 
\textit{Boundary conditions for integrable quantum systems}, 
J. Phys. A: Math. Gen. \textbf{21}, 2375--2389 (1988).

\bibitem{Taka71b}
M. Takahashi, \textit{One-dimensional {Heisenberg} model at finite
  temperature}, Prog. Theor. Phys. \textbf{46}, 401--415 (1971).

 
\bibitem{tafhlt2008} 
S. Trebst, E. Ardonne, A. Feiguin, D.A. Huse, A.W.W. Ludwig and M. Troyer, 
\textit{Collective states of interacting Fibonacci anyons}, 
Phys. Rev. Lett. \textbf{101}, 050401 (2008).  

\bibitem{ttwl08} 
S. Trebst, M. Troyer, Z. Wang, A.W.W. Ludwig, 
\textit{A short introduction to Fibonacci anyon models},  
Prog. Theor. Phys. Supp. \textbf{176}, 384--407 (2008). 

\end{thebibliography}

\end{document}